\documentclass[12pt]{article}
\usepackage{amssymb}

\renewcommand{\d}{\delta}

\newcommand{\D}{\Delta}

\def\be{\begin{equation}}
\def\ee{\end{equation}}
\def\bea{\begin{eqnarray}}
\def\eea{\end{eqnarray}}
\def\ba{\begin{array}}
\def\ea{\end{array}}
\def\bi{\begin{itemize}}
\def\ei{\end{itemize}}

\catcode`\@=11
\def\@citex[#1]#2{%
\if@filesw \immediate \write \@auxout {\string \citation {#2}}\fi
\@tempcntb\m@ne \let\@h@ld\relax \def\@citea{}%
\@cite{%
  \@for \@citeb:=#2\do {%
    \@ifundefined {b@\@citeb}%
      {\@h@ld\@citea\@tempcntb\m@ne{\bf ?}%
      \@warning {Citation `\@citeb ' on page \thepage \space undefined}}%
      {\@tempcnta\@tempcntb \advance\@tempcnta\@ne%
      \@tempcntb\number\csname b@\@citeb \endcsname \relax%
      \ifnum\@tempcnta=\@tempcntb 
        \ifx\@h@ld\relax%
          \edef \@h@ld{\@citea\csname b@\@citeb\endcsname}%
        \else%
          \edef\@h@ld{\ifmmode{-}\else--\fi\csname b@\@citeb\endcsname}%
        \fi%
      \else
        \@h@ld\@citea\csname b@\@citeb \endcsname%
        \let\@h@ld\relax%
      \fi}%
    \def\@citea{,\penalty\@highpenalty\,}%
  }\@h@ld
}{#1}}

\def\@citeb#1#2{{[#1]\if@tempswa , #2\fi}}
\def\@citeu#1#2{{$^{#1}$\if@tempswa , #2\fi }}
\def\@citep#1#2{{#1\if@tempswa , #2\fi}}

\def\bcites{         
        \catcode`\@=11
        \let\@cite=\@citeb
        \catcode`\@=12
}

\def\upcites{         
        \catcode`\@=11
        \let\@cite=\@citeu
        \catcode`\@=12
}

\def\plaincites{      
        \catcode`\@=11
        \let\@cite=\@citep
        \catcode`\@=12
}

\newcount\hour
\newcount\minute
\newtoks\amorpm
\hour=\time\divide\hour by 60
\minute=\time{\multiply\hour by 60 \global\advance\minute by-\hour}
\edef\standardtime{{\ifnum\hour<12 \global\amorpm={am}%
        \else\global\amorpm={pm}\advance\hour by-12 \fi
        \ifnum\hour=0 \hour=12 \fi
        \number\hour:\ifnum\minute<10 0\fi\number\minute\the\amorpm}}
\edef\militarytime{\number\hour:\ifnum\minute<10 0\fi\number\minute}

\def\draftlabel#1{{\@bsphack\if@filesw {\let\thepage\relax
   \xdef\@gtempa{\write\@auxout{\string
      \newlabel{#1}{{\@currentlabel}{\thepage}}}}}\@gtempa
   \if@nobreak \ifvmode\nobreak\fi\fi\fi\@esphack}
        \gdef\@eqnlabel{#1}}
\def\@eqnlabel{}
\def\@vacuum{}
\def\marginnote#1{}
\def\draftmarginnote#1{\marginpar{\raggedright\scriptsize\tt#1}}
\def\draft{
        \pagestyle{plain}
        \overfullrule=2pt
        \oddsidemargin -.5truein
        \def\@oddhead{\sl \phantom{\today\quad\militarytime} \hfil
        \smash{\Large\sl DRAFT} \hfil \today\quad\militarytime}
        \let\@evenhead\@oddhead
        \let\label=\draftlabel
        \let\marginnote=\draftmarginnote
        \def\ps@empty{\let\@mkboth\@gobbletwo
        \def\@oddfoot{\hfil \smash{\Large\sl DRAFT} \hfil}
        \let\@evenfoot\@oddhead}
        \def\@eqnnum{(\theequation)\rlap{\kern\marginparsep\tt\@eqnlabel}%
        \global\let\@eqnlabel\@vacuum}  }

\begin{document}



\hfill UTHET-01-0501

\vspace{-0.2cm}

\begin{center}
\Large
{ \bf Entropy bounds in cosmology and conformal field theories\footnote{Presented at ``Frontiers
in Contemporary Physics II," Vanderbilt University, March 2001}}
\normalsize

\vspace{0.6cm}

\baselineskip=14pt
{\bf G. Siopsis}\footnote{gsiopsis@utk.edu}
\footnote{Research supported in part by the US Department of Energy under grant
DE--FG05--91ER40627.}\\ Department of Physics
and Astronomy,
The University of Tennessee, Knoxville,\\
TN 37996 - 1200, USA.
 \end{center}

\vspace{0.2cm}
\baselineskip=12pt

{\bf Abstract:}
{\em We study the ratio of the entropy to the total energy in conformal field
theories at
finite temperature. For the free field realizations of ${\cal N}=4$
super Yang-Mills theory in $D=4$ and the $(2,0)$ tensor
multiplet in $D=6$, the ratio is bounded from above. The corresponding bounds
are  less stringent  
than the recently proposed Verlinde
bound. For
strongly coupled CFTs with AdS duals, we show that the ratio obeys the
Verlinde bound even in the presence of rotation. For such CFTs, we
point out an
intriguing resemblance in their thermodynamic formulas with 
the corresponding ones of two-dimensional
CFTs.\footnote{The discussion is based on ref.~\cite{ego}}
}


\normalsize
\baselineskip=14pt\parskip=10pt
\section{Introduction}
The Bekenstein bound \cite{Bekenstein} for the ratio of the entropy
$S$ to the total energy $E$ of a closed physical system that fits in
a sphere in three spatial dimensions reads ~\cite{Veneziano}
\be
\label{Bbound}
\frac{S}{2\pi RE} \leq 1,
\ee
where $R$ denotes the radius of the sphere.
Despite many efforts, the microscopic origin of the bound  
remains elusive. A recent interesting development is Verlinde's
observation \cite{Verlinde} that
CFTs possessing AdS duals satisfy a version of the bound
(\ref{Bbound}). One firstly observes that for general CFTs on
${\mathbb R} \times S^{D-1}$, 
with the radius of $S^{D-1}$ being $R$, the product $ER$ is
independent of the total spatial volume $V$. If one defines
the sub-extensive part $E_C$ of the total energy through the scaling
property $E_C(\lambda S,\lambda V)=\lambda^{1-\frac{2}{D-1}}E_C(S,V)$
and $E=E_{ext}+\frac{1}{2}E_C$, it follows that
\be
\label{subext}
E_C = DE -(D-1)TS\,.
\ee
The observation of Verlinde is that for strongly coupled CFTs with AdS
duals the entropy is given by a generalized Cardy formula
\be
\label{verlentr}
S=\frac{2\pi R}{D-1}\sqrt{E_C(2E-E_C)}\,.
\ee
To show this, one employs the results for the entropy and total
energy of the corresponding $D$-dimensional CFT that fits into a
$(D-1)$-dimensional sphere at finite temperature \cite{Witten1}.
From (\ref{verlentr}) one obtains a bound similar to (\ref{Bbound}),
namely\footnote{Henceforth we call (\ref{BVerl}) the Verlinde CFT
bound to avoid confusion with the cosmological entropy bound suggested
by Verlinde also in \cite{Verlinde}.}
\be
\label{BVerl}
\frac{S}{2\pi R E}\leq \frac{1}{D-1}\,.
\ee

In view of the above developments, a natural question arising is
whether there exists a
microscopic derivation of Verlinde's formula (\ref{verlentr}) within
the thermodynamics of CFTs. This question could be checked in the
context of CFTs whose  
microscopic thermodynamics is well understood, such as free CFTs on
${\mathbb R} \times S^{D-1}$~\cite{KL}.

Here, we further analyze the results in
\cite{KL} for free CFTs in dimensions $D=4,6$. We find that for the
specific cases of ${\cal N}=4$ $U(N)$ SYM theory in $D=4$ and the $(2,0)$ tensor
multiplet in $D=6$, the ratio of the entropy to the total energy is
bounded from 
above, however the corresponding bounds are less stringent than
(\ref{BVerl}). Although bounds for the
ratio of the entropy to the total 
energy seem to arise quite generically in CFTs, their exact values
depend on the details of 
the underlying CFT, {\em e.g.,} it seems
that the bounds become more stringent as one goes from weak to strong
coupling.

Next we turn our attention to strongly coupled CFTs with AdS duals. We
show that  the Verlinde
formula (\ref{verlentr}) remains  valid also in the case of
strongly coupled CFTs in a rotating Einstein universe. We then point out an intriguing
resemblance of the formulas of $(D+1)$-dimensional AdS black hole
thermodynamics to corresponding formulas in the thermodynamics of two-dimensional
CFTs. Particularly interesting is the fact that the entropy of the
black hole resembles the $C$-function of a two-dimensional
system.

\section{CFTs at weak coupling}


In this section, we discuss the thermodynamics of conformal field theories.
For a general statistical mechanical system, one defines the partition
function (we put $k=\hbar=1$)
\be\label{eq2}
Z = \sum_E \rho (E) e^{-E/T}\,,
\ee
where $\rho (E)$ is the number of states with energy $E$. In general,
(\ref{eq2}) can be evaluated
using a saddle point approximation. The exponent is stationary when
\be\label{eqstar}
dS = {dE\over T}\,,\quad S = \ln\rho\,,
\ee
where $S$ is the entropy of the system. This approximation is valid for a large
number of degrees of freedom, i.e. if the underlying theory is a CFT
with a large central charge. The free energy
\be\label{eq3}
F = -T\ln Z\,,
\ee
at the saddle point is the exponent in~(\ref{eq2}),
\be\label{eq4}
F = E-TS\,,
\ee
and on account of~(\ref{eqstar}), the entropy is given by
\be\label{eq5}
S = - {\partial F\over\partial T}\,.
\ee
As an application, consider $C$ free massless bosons living on a three-dimensional
spatial sphere of radius $R$ at finite temperature. The energy levels and
corresponding degeneracies of the 
various modes are \cite{KL}
\be
E_n = {n\over R} \,,\quad d_n = n^2\,.
\ee
Therefore, the partition function reads
\be
Z_B^{(4)} = \prod_{n=1}^\infty (1-q^n)^{-Cn^2} \,,\quad
q= e^{-2\pi\delta} \,,\quad \delta = {1\over 2\pi RT}\,.
\ee
To cast this into the form~(\ref{eq2}), we exploit the modular properties of
the partition function $Z_B^{(4)}$. Using a Mellin transform, one
obtains
\be
Z_B^{(4)} = e^{{\pi C\over 360}\; \delta^{-3}}\;
e^{{\pi C\over 120}\; \delta} \; {\cal Z}\,,
\ee
where ${\cal Z}$ is a slowly varying function (approximately constant)
near the saddle point. Therefore, the free energy~(\ref{eq3}) in the
saddle point approximation is
\be
\label{F4}
- F_B^{(4)} R = {C\over 240} \left( {1\over 3} \delta^{-4} +1\right)\,,
\ee
where we multiplied by the negative radius $-R$ for convenience. The entropy
and energy of the system are easily deduced from the thermodynamical
relations~(\ref{eq4}) and (\ref{eq5}),
\be
\label{S4}
S = {\pi C\over 90}\; \delta^{-3}\,,\quad
E = {C\over 240}\; \left( \delta^{-4} -1\right)\,.
\ee
For comparison, in two dimensions, the partition function of $C$ free bosons is
\be
Z_B^{(2)} = \prod_{n=1}^\infty (1-q^n)^{-C}\,,
\ee
leading to the free energy
\be
\label{FB2}
- F_B^{(2)} R = {C\over 24} \; \left( \delta^{-2} - 1\right)\,,
\ee
while the entropy and energy read, respectively
\be\label{eq7}
S = {\pi C\over 6}\; \delta^{-1}\,,\quad
ER = {C\over 24} \; \left( \delta^{-2} + 1\right)\,.
\ee
We should point out that in this case there is a contribution from the zero
modes of the form $\delta^{C/2}$. This is a slowly varying function and does
not contribute to the rapidly varying exponentials that comprise the free
energy at the saddle point. The contribution of the zero modes becomes
significant when the saddle-point approximation breaks down, for a small
central charge.\footnote{We thank D.~Kutasov for pointing out this to us.}
Here, we are interested in the large $C$ limit, so such contributions will be
ignored.

Eq.~(\ref{eq7}) implies the Cardy formula \cite{Cardy3}
\be
S = 2\pi \sqrt{{C\over 6}\; \left( E - {C\over 24} \right)}\,,
\ee
and the Bekenstein bound (\ref{Bbound}) for the ratio
\be
{S\over 2\pi ER} = {2\delta\over 1 +\delta^2} \le 1\,.
\ee
The above results also hold for fermions, because the free energy for a
fermion is $F_F^{(2)} = {1\over 2}\; F_B^{(2)}$.

Returning to four dimensions, we note from (\ref{S4}) that there is a
transition point at $\d=1$ where 
$ER=0$. At that point the ratio $S/E$ diverges
because the entropy remains finite. Thus, it seems as if the ratio of
the entropy to the total energy is not bounded for free
bosons. Similar results hold for Weyl fermions and vector
bosons. However, in systems with diverse mode species there is a
chance that the above ratio is bounded. Consider the free
energy of a system of $N_B$ bosons, $N_F$ Weyl fermions and $N_V$ vectors
that reads \cite{KL}
\be
\label{FR4}
-FR = a_4 \delta^{-4} + a_2 \delta^{-2} + a_0\,,
\ee
where
\be
a_4 = {\textstyle{1\over 720}} ( N_B + {\textstyle{7\over 4}}\; N_F +2N_V)
\;,\quad
a_2 = -{\textstyle{1\over 24}} ( {\textstyle{1\over 4}}\; N_F + 2N_V)
\;,\quad a_0 = {\textstyle{1\over 240}} ( N_B +{\textstyle{17\over 4}}\; N_F
+ 22N_V) \,,\ee
satisfying the constraint $3a_4 = a_2+a_0$. The entropy and energy are,
respectively,
\be
\label{SER4}
S = 2\pi\; (4a_4\delta^{-3} + 2a_2\delta^{-1})\;,\quad
ER = 3a_4\delta^{-4}+a_2\delta^{-2} - a_0\,.
\ee
The Bekenstein-Verlinde ratio is
\be
\label{SE4}
{S\over 2\pi ER} = \delta\; {4a_4+2a_2\delta^2\over 3a_4+a_2\delta^2
-a_0\delta^4} = {2\delta\over 1+\delta^2} \; {2a_4+ a_2\delta^2\over
3a_4-a_0\delta^2}\,.
\ee
Remarkably, for the ${\cal N}=4$ SYM model, we have $a_2 = -6a_4$,
which implies
\be
{S\over 2\pi ER} = {2\over 3}\; {2\delta\over 1+\delta^2}\,.\label{SE44}
\ee
One might now think that (\ref{SE44}) generally implies the bound
$S/(2\pi ER) \leq 2/3$. However, we have to keep in mind that
(\ref{FR4}) and (\ref{SER4}) are high temperature expansions and
should not be trusted for large-$\d$. Starting from high
temperatures (small-$\d$), there is a critical point at which both $S$
and $ER$ vanish for
\be
\delta_c^2 = - {2a_4\over a_2} = {1\over 3}\,.
\ee
We should not expect that (\ref{SE4}) makes sense for
$\d\ge\d_c$. Nevertheless, for $\delta\le \delta_c$, we obtain the bound
\be
{S\over 2\pi ER} \le {\sqrt 3\over 3}\,,
\ee
which is weaker than the Verlinde CFT bound (\ref{BVerl}) $S/(2\pi ER)
\le 1/3$. 

It is perhaps worth mentioning that if one imposes periodic boundary
conditions on the gaugino, as suggested by Tseytlin \cite{GKP} to
account for the 
disagreement on the number of degrees of freedom between the weak and strong
coupling regimes, the above results still hold. Indeed, the partition function
for a gaugino with periodic boundary conditions is
\be
\widetilde Z_F = \prod_{n=0}^\infty (1-q^{n+1/2})^{2n(n+1)}\,,
\ee
which leads to the free energy for $\widetilde N_F$ gauginos,
\be
\widetilde FR = \widetilde N_F (\widetilde a_4 \delta^{-4} +
\widetilde a_2\delta^{-2} + \widetilde a_0)\,,
\ee
where
\be
\widetilde a_4 = - {1\over 360} \,,\quad
\widetilde a_2 = {1\over 48} \,,\quad
\widetilde a_0 = - {7\over 240}\,.
\ee
For the ${\cal N}=4$ SYM model with such gauginos, the coefficients of the
free energy (\ref{FR4}) still satisfy $3a_4' = a_2'+a_0'$ and the ratio
$a_2'/a_4'$ is unchanged 
\be
{a_2'\over a_4'} = -30 \; {{1\over 4} \; N_F - {1\over 2}\; \widetilde N_F
+ 2N_V\over N_B + {7\over 4}\; N_F -2\widetilde N_F +2N_V} = -6\,.
\ee
Thus, the only effect of imposing periodic
boundary conditions on the gauginos is the reduction of the free energy by
an overall factor of $a_4'/a_4 = 3/4$.

Next we turn to the $(2,0)$ tensor multiplet in $D=6$. The free energy is \cite{KL}
\be
-FR = a_6\delta^{-6} + a_4 \delta^{-4} + a_2 \delta^{-2} + a_0\,.
\ee
The entropy is given by 
\be
{S\over 2\pi} = TSR\delta = - T\; {\partial (FR)\over\partial T}\delta =
6a_6\delta^{-5} + 4a_4\delta^{-3} + 2a_2\delta^{-1}\,,
\ee
and the energy by
\be
ER= 5a_6\delta^{-6} + 3a_4\delta^{-4} + a_2\delta^{-2} -a_0\,.
\ee
Considering the same
ratio as before we obtain
\be
{S\over 2\pi RE} = \delta\; {6a_6+4a_4\delta^2+2a_2\delta^4
\over 5a_6+3a_4\delta^2 + a_2\delta^4
- a_0\delta^6}\,.
\ee
The coefficients are related through \cite{KL}
\be
5a_6-3a_4+a_2+a_0=0\,,
\ee
and we have $a_6, a_2>0$, $a_4, a_0<0$\,.
The denominator can be factorized into
\be
5a_6+3a_4\delta^2 + a_2\delta^4
- a_0\delta^6 = (1+\delta^2)(5a_6+(a_2+a_0)\delta^2-a_0\delta^4)\,.
\ee
Using the explicit values $a_6 = 1/5$, $a_4=-5/3$, $a_2=19$,
$a_0 = -25$ we obtain
\be
{S\over 2\pi ER} = {2\delta\over 1+\delta^2}\; {{3\over 5}-{10\over 3}\delta^2+19\delta^4
\over 1-6\delta^2 + 25\delta^4}\,.
\ee
This is a well-behaved function of $\delta$, which has a maximum
of $0.824$. We therefore
conclude that
\be
{S\over 2\pi ER} \le 0.824\,,
\ee
which is less stringent than the Verlinde CFT bound (\ref{BVerl}).

\section{CFTs at strong coupling}


In this section we turn our attention to strongly coupled CFTs in
$D$-dimensions, possessing AdS duals. The thermodynamics of such
theories follows quite generally from the thermodynamics of
$(D+1)$-dimensional AdS black holes, through holography. 
We consider the rotating Kerr-AdS (KAdS) black hole in $(D+1)$-dimensions given
by \footnote{For simplicity we restrict ourselves to the
case of only one rotation parameter.} \cite{hawk98}
\begin{eqnarray}
ds^2 &=& -\frac{\Delta_r}{\rho^2}[dt - \frac{a}{\Xi}\sin^2\theta d\phi]^2
         + \frac{\rho^2}{\Delta_r}dr^2 + \frac{\rho^2}{\Delta_{\theta}}
         d\theta^2 \nonumber \\
     & & + \frac{\Delta_{\theta}\sin^2\theta}{\rho^2}[adt - \frac{r^2+a^2}{\Xi}
         d\phi]^2 + r^2\cos^2\theta d\Omega^2_{D-3}, \label{KAdS}
\end{eqnarray}
where $d\Omega^2_{D-3}$ denotes the standard metric on the unit $S^{D-3}$ and
\begin{eqnarray}
\Delta_r &=& (r^2 + a^2)\left(1 + \frac{r^2}{R^2}\right) - 2Mr^{4-D},
             \nonumber \\
\Delta_{\theta} &=& 1 - \frac{a^2}{R^2}\cos^2\theta, \\
\Xi &=& 1 - \frac{a^2}{R^2}, \nonumber \\
\rho^2 &=& r^2 + a^2\cos^2\theta.
\end{eqnarray}
The inverse temperature, free energy, entropy, energy and angular momentum
read \cite{hawk98,cald99}
\begin{eqnarray}
\beta &=& \frac{4\pi(r_+^2+a^2)}{(D-2)\left(1+\frac{a^2}{R^2}\right)r_+
          + \frac{Dr_+^3}{R^2} + \frac{(D-4)a^2}{r_+}}, \label{b}  \\
F &=& -\frac{V_{D-1}}{16\pi G_{D+1}\Xi}r_+^{D-4}(r_+^2+a^2)\left(\frac{r_+^2}
      {R^2} - 1\right), \label{FE}\\
S &=& \frac{V_{D-1}}{4G_{D+1}\Xi}r_+^{D-3}(r_+^2+a^2), \label{S} \\
E &=& \frac{(D-1)V_{D-1}}{16\pi G_{D+1}\Xi}r_+^{D-4}(r_+^2+a^2)
      \left(\frac{r_+^2}{R^2} + 1\right), \label{E} \\
J &=& \frac{aV_{D-1}}{8\pi G_{D+1}\Xi^2}r_+^{D-4}(r_+^2+a^2)
      \left(\frac{r_+^2}{R^2} + 1\right), \label{J}
\end{eqnarray}
where $G_{D+1}$ is Newton's constant, $V_{D-1}$ denotes the volume
of the unit $S^{D-1}$, $r_+$ (the horizon radial coordinate)
is the largest root of $\Delta_r=0$,
and the rotation parameter $a$ is restricted to the range $0 \le a < R$.
According to the AdS/CFT duality conjecture \cite{Witten1}, the above
thermodynamical 
quantities are associated to a strongly coupled $D$-dimensional CFT
residing on the conformal boundary of the spacetime (\ref{KAdS}),
i.~e.~on a rotating Einstein universe.

Defining $\D = R/r_+$, and the Bekenstein entropy $S_B = 2\pi ER/(D-1)$,
we obtain from (\ref{S}) and (\ref{E})
\begin{equation}
\frac{S}{S_B} = \frac{2\D}{1+\D^2} \le 1.
\end{equation}
The Bekenstein bound is saturated for $\D = 1$, i.~e.~at the Hawking-Page
transition point \cite{HP}, where the free energy becomes zero.
We further note that we can write
\begin{equation}
2ER = \frac{D-1}{2\pi}S\frac{R}{r_+}[\D^{-2} + 1]\,, \label{2ER}
\end{equation}
which, for arbitrary $D$, is exactly the behavior of a two-dimensional
CFT (\ref{eq7}) with
characteristic scale $R$, temperature $\tilde T = 1/(2\pi R\D) =
r_+/(2\pi R^2)$,
and central charge proportional to $SR/r_+$.
This resemblance motivates us 
to define the Casimir energy as the sub-extensive part of (\ref{2ER}),
i.~e.
\begin{equation}
E_C = \frac{(D-1)}{2\pi}\frac{S}{r_+} = \frac{D-1}{2\pi}\frac{V_{D-1}}
      {4G_{D+1}\Xi}r_+^{D-4}(r_+^2+a^2). \label{Casimir}
\end{equation}
Note that in the non-rotating case $a=0$, (\ref{Casimir}) coincides
with the expression \footnote{The expressions
in \cite{Verlinde} are related to ours by $L=R^2/r_+$ and $\frac{c}{12}
\frac{V}{R^{2D-2}} = \frac{V_{D-1}}{4G_{D+1}}$.} given in \cite{Verlinde}.

One now easily verifies that the quantities (\ref{S}), (\ref{E}) and
(\ref{Casimir}) satisfy exactly the Verlinde formula (\ref{verlentr}).
We can also define the "Casimir entropy" \cite{Verlinde} by
\begin{equation}
S_C = \frac{2\pi}{D-1}E_CR = S\frac{R}{r_+}\,, \label{SC}
\end{equation}
which allows to write the free energy as
\begin{equation}
-FR = \frac{S_C}{4\pi}[\D^{-2} - 1]. \label{FRSC}
\end{equation}
Comparing (\ref{FRSC}) with the corresponding relation of a two-dimensional
bosonic CFT (\ref{FB2}), we see that the Casimir entropy $S_C$ is
essentially proportional to the central charge \cite{Verlinde},
or equivalently to the number of degrees of
freedom coupled at the critical point.

Within such an
interpretation for $S_C$ we can now see that the temperature
$\tilde{T}=r_+/(2\pi R^2)$ makes thermodynamic sense as a temperature
of a two-dimensional system. Considering for simplicity the case when
$a=0$ and constant volume, the second law of black hole
mechanics reads
\be
dE(S,N) = TdS + \mu dN\,,
\ee
where by virtue of (\ref{SC}) and (\ref{S}) we defined the number 
 of degrees of freedom (generalized central charge) as
\be
\label{N}
N = {V_{D-1}R^{D-1}\over 16\pi G_{D+1}}={S_C\over 4\pi}
    \left({R \over r_+}\right)^{D-2}\,,
\ee
and $\mu$ is the chemical potential.
After some algebra, we find
\be
\mu = - \left({r_+ \over R}\right)^{D-2} \left( {r_+^2\over R^2} -1\right)
      \frac{1}{R}\,.
\ee
The free energy (\ref{FRSC}) can then be written simply as 
\be
F = \mu N\,.
\ee
Furthermore, for $S_C =$~const., we have 
\be
d\tilde{E} = \tilde{T}\; dS\,,
\ee
where
\be
\tilde{E} = E\frac{1}{D-1}\quad,\quad \tilde{T} = {1\over D-1}\;
\left( T + \mu {dN\over dS} \right) = {r_+\over 2\pi R^2}\,.
\ee
Notice that $\tilde{E} = E$ and $\tilde{T}=T$ for $D=2$, as expected.

Remarkably enough, these results generalize to the $a\ne 0$ case. Even
with the addition of one more potential and the attendant generalization of
the second law of black hole mechanics to
\be
dE = TdS + \mu dN + \Omega dJ\,,
\ee
the condition $dS_C=0$ still describes a two-dimensional system.

Finally, as in \cite{Verlinde} we define the "Bekenstein-Hawking energy"
$E_{BH}$ as the energy for which the black hole entropy $S$ (\ref{S}) and
the Bekenstein entropy $S_B$ are equal, $2\pi E_{BH}R/(D-1) = S$,
yielding
\begin{equation}
E_{BH} = E_C\frac{r_+}{R}.
\end{equation}
One checks that $E_{BH} \le E$. Furthermore, because
we are above the Hawking-Page transition point, we have
$r_+ \ge R$, and therefore
\begin{equation}
E_C \le E_{BH} \le E, \qquad S_C \le S \le S_B,
\end{equation}
where equality holds when the HP phase transition is reached.
As the entropy $S$ is a monotonically increasing function of $E_C$
(or, equivalently, of $r_+$), the maximum entropy is reached when
$E_C = E_{BH}$, i.~e.~at the HP phase transition $r_+ = R$.
It is quite interesting to observe that at this point the central charge
$c/12 = S_C/(2\pi)$ takes e.~g.~for $D=4$, ${\cal N}=4$ $U(N)$ SYM
theory the value\footnote{Here we used the AdS/CFT dictionary 
$N^2 = \frac{\pi R^3}{2G_5}$.} $c=6N^2$. This is exactly the central charge
of a two-dimensional free CFT containing the $6N^2$ scalars of
$D=4$, ${\cal N}=4$ SYM.

\section{Cosmological implications}

In this section, we discuss the implications of our results for cosmology. As in in
Ref.~\cite{Verlinde}, consider a radiation
dominated closed Friedman-Robertson-Walker
(FRW) universe. The FRW metric takes the form
\begin{equation}
ds^2 = -d\tau^2 + {\cal R}^2(\tau)d\Omega^2_{D-1}, \label{FRW}
\end{equation}
where ${\cal R}(\tau)$ represents the radius of the universe at
a given time $\tau$. Note that the metric (\ref{FRW}) is conformally
equivalent to
\begin{equation}
d\tilde s^2 = -dt^2 + R^2d\Omega^2_{D-1}, \label{confmetric}
\end{equation}
where $dt = R\,d\tau/{\cal R}(\tau)$. If the radiation is described
by a CFT, one can equally well use (\ref{confmetric}) instead of (\ref{FRW}).
If, in addition, this CFT admits an AdS dual, it can be described by
a Schwarz\-schild-AdS black hole at some temperature $T$, because
(\ref{confmetric}) is precisely the metric on the conformal boundary
of (\ref{KAdS}) for $a=0$.
The observations made by Verlinde \cite{Verlinde} concerning
entropy, energy and temperature bounds in a radiation dominated
universe then fit nicely into this AdS black hole description.
In particular, the universe is weakly (strongly) self-gravitating
if $H{\cal R} \le 1$ ($H{\cal R} \ge 1$), where $H=\dot{{\cal R}}/{\cal R}$
denotes the Hubble constant, and the dot refers to differentiation
with respect to $\tau$. One has $H{\cal R} = 1$
iff the Bekenstein-Hawking entropy $S$ equals the Bekenstein
entropy $S_B$. We saw above that this happens precisely at the
HP transition point $r_+ = R$, so the borderline between the
weakly and strongly self-gravitating regime is the Hawking-Page
phase transition temperature $T_{HP} = (D-1)/2\pi R$. This identification
makes indeed sense, because below $T_{HP}$ (weakly gravitating)
one has AdS space filled with thermal radiation which collapses
above $T_{HP}$ ($r_+ \ge R$, strongly gravitating) to form a black
hole. Furthermore, in \cite{Verlinde} a limiting temperature was
found for the early universe,
\begin{equation}
T \ge T_H = -\frac{\dot{H}}{2\pi H} \qquad \mbox{for} \quad H{\cal R} \ge 1.
\end{equation}
We conclude therefore that Verlinde's limiting temperature $T_H$
corresponds to the temperature $T_{HP}$ where the HP phase transition
takes place.

\section{Conclusions}

Concerning further developments of our ideas, it might be interesting
to study our {\it generalized central 
charge} defined in (\ref{N}). We discussed in the text that this
quantity intriguingly resembles a standard 
two-dimensional central charge. Such an interpretation leads to the
conjecture that there might exist a two-dimensional CFT model whose
dynamics in the presence of irrelevant operators underlies 
the dynamics of the $D$-dimensional CFTs possessing AdS duals. Such a
conjecture might explain the fact that the latter theories
share unexpectedly many of the properties of two-dimensional CFTs
\cite{Anselmi}.

It would also be interesting to understand what
happens if the sub-extensive part of the total energy becomes
negative. This is e.~g.~the case for hyperbolic AdS black holes.
As $E_C$ corresponds somehow to a central charge,
this would indicate that the underlying CFT is non-unitary or that the
theory has a thermally unstable ground state.

\end{document}